\documentclass{article}
\usepackage{balance}

\usepackage{microtype}
\usepackage{graphicx}
\usepackage{subcaption}
\usepackage{booktabs} % for professional tables

\usepackage{hyperref}

\usepackage[preprint]{icml2026}

\usepackage{amsmath, amssymb} 
\usepackage{bm}
\usepackage{textcomp}
\usepackage{gensymb}
\usepackage{pifont}
\usepackage{enumitem}
\usepackage[T1]{fontenc}
\usepackage{multirow}
\usepackage{makecell}
\usepackage{xspace}
\usepackage{url}
\usepackage{enumerate}

\usepackage{algorithm}
\usepackage{algorithmic}

\usepackage[table]{xcolor}
\usepackage{siunitx}
\definecolor{lightblue}{HTML}{E0F2F7}

\makeatletter

\DeclareRobustCommand\onedot{\futurelet\@let@token\@onedot}
\def\@onedot{\ifx\@let@token.\else.\null\fi\xspace}
\def\eg{\emph{e.g}\onedot} 
\def\ie{\emph{i.e}\onedot}

\makeatother

\DeclareUnicodeCharacter{2212}{-}

\icmltitlerunning{EEG-Based Brain-LLM Interface for Human Preference Aligned Generation}

\begin{document}
\raggedbottom

\twocolumn[
  \icmltitle{EEG-Based Brain-LLM Interface for Human Preference Aligned Generation}

  \begin{icmlauthorlist}
    \icmlauthor{Junzi Zhang}{sdu}
    \icmlauthor{Jianing Shen}{sdu}
    \icmlauthor{Weijie Tu}{anu}\hspace{5.5in}
    \icmlauthor{Yi Zhang}{sdu}
    \icmlauthor{Hailin Zhang}{sdu}
    \icmlauthor{Tom Gedeon}{curtin}
    \icmlauthor{Bin Jiang}{sdu}
    \icmlauthor{Yue Yao$^\dagger$}{sdu}
  \end{icmlauthorlist}

  \icmlaffiliation{sdu}{Shandong University, China}
  \icmlaffiliation{anu}{Australian National University, Australia}
  \icmlaffiliation{curtin}{Curtin University, Australia}

  % 这里的通讯作者信息会自动生成在首页下方的脚注
  \icmlcorrespondingauthor{Yue Yao}{yaoyorke@gmail.com}

  \icmlkeywords{Brain-Computer Interface, LLM, EEG, Alignment}

  \vskip 0.3in
]

% 打印单位信息和通讯作者脚注
\printAffiliationsAndNotice{}

\begin{abstract}

% Large language models (LLMs) have become central to human–machine interaction. 
% However, they often generate responses that only partially reflect user intent, leading to repeated prompting and inefficiency. Current alignment techniques rely on explicit linguistic feedback, which captures deliberate judgments but often overlooks users’ implicit cognitive and affective responses during interaction.

Large language models (LLMs) are becoming an increasingly important component of human–computer interaction, enabling users to coordinate a wide range of intelligent agents through natural language. While language-based interfaces are powerful and flexible, they implicitly assume that users can reliably produce explicit linguistic input, an assumption that may not hold for users with speech or motor impairments, \eg, Amyotrophic Lateral Sclerosis (ALS). In this work, we investigate whether neural signals can be used as an alternative input to LLMs, particularly to support those socially marginalized or underserved users. We build a simple brain-LLM interface, which uses EEG signals to guide image generation models at test time. 
Specifically, we first train a classifier to estimate user satisfaction from EEG signals. Its predictions are then incorporated into a test-time scaling (TTS) framework that dynamically adapts model inference using neural feedback collected during user evaluation. 
% When EEG activity indicates possible dissatisfaction, the model performs an additional refinement step; otherwise, it returns the response directly to preserve efficiency. 
% We evaluate the proposed EEG-guided adaptation in a human study with participants using $64$-channel EEG recorded across multiple sessions. 
The experiments show that EEG can predict user satisfaction, suggesting that neural activity carries information on real-time preference inference. 
% The proposed pipeline also improves the alignment between model outputs and user preferences compared with a baseline without EEG feedback. 
These findings provide a first step toward integrating neural feedback into adaptive language-model inference, and hopefully open up new possibilities for future research on adaptive LLM interaction.
\end{abstract}

% =======================================================
% MAIN CONTENT START
% =======================================================

\section{Introduction}
\label{sec:intro}

Large language models are becoming an increasingly important component of modern human–computer interaction (HCI)~\cite{radford2019rewon, brown2020language, openai2023gpt4}, supporting applications in content generation~\cite{ouyang2022training, raffel2020exploring}, information retrieval~\cite{lewis2020retrieval, guu2020retrieval}, and user engagement~\cite{thoppilan2022lamda, zhang2020dialogpt}. By producing coherent and context-aware text, LLMs assist with tasks such as writing support~\cite{lee2022coauthor, ouyang2022training}, question answering~\cite{lewis2020retrieval, wei2022chain}, and data synthesis~\cite{zhang2020pegasus, wei2021finetuned}. Their use has also expanded to multimodal systems, where textual prompts guide models to generate visual~\cite{ramesh2022hierarchical, saharia2022photorealistic} or auditory~\cite{borsos2023audiolm, wang2023neural} content, linking linguistic input with other representational modalities.

\begin{figure}[t]
    \centering
    % Ensure the filename matches what you uploaded to Overleaf
    \includegraphics[width=\linewidth]{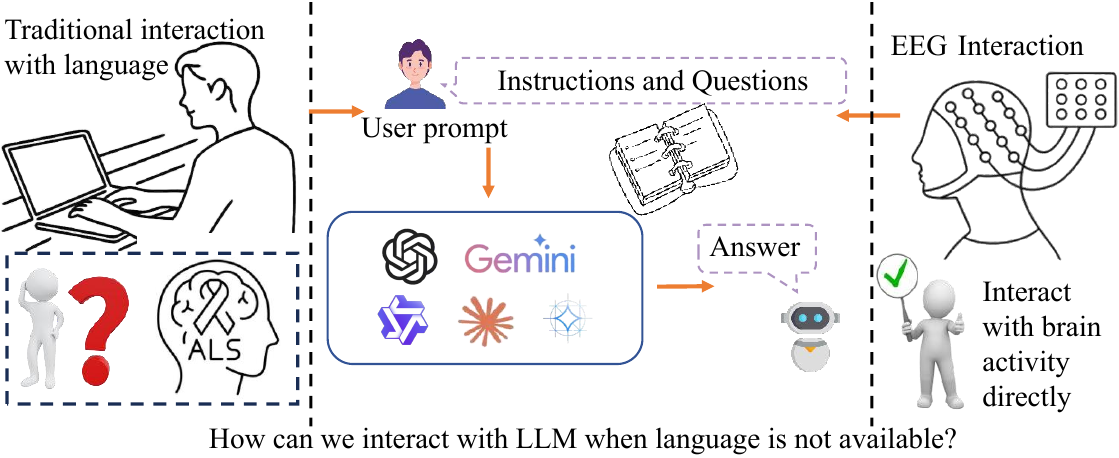} 
    \caption{\textbf{Motivation for EEG-based interface for LLM.} Existing large language models rely mainly on explicit linguistic feedback such as typed instructions or corrections. However, there may exist a group of users with speech or motor impairments (\eg, ALS) thus language may not be available. For those users, we investigate whether neural signals can be used as an alternative input to LLMs.}
    \label{fig:intro}
    \vspace{-1em}
\end{figure}

Despite these advances, LLM-based interaction may not equally serve all users, as it fundamentally relies on the availability of explicit linguistic input. This assumption can break down in scenarios where users have limited ability to produce speech or text, or when preferences are difficult to articulate through language alone. As illustrated in Figure~\ref{fig:intro}, when a user is influenced by ALS, such limitations reflect constraints on the interaction channel rather than on users’ underlying cognitive intent: in many cases, individuals may retain intact cognitive function and clear preferences, while their ability to express them through language is restricted. This mismatch motivates the exploration of alternative, non-linguistic feedback signals that can complement language-based interfaces.

% LLMs may not well serve all groups of people as it requires explicit language as input. Shown in Figure~\ref{fig:intro}, their outputs may appear fluent yet diverge from what users actually expect or prefer~\cite{ji2023survey}. Textual prompts effectively convey explicit goals~\cite{hancock2019feedback, kocielnik2018affective}, but implicit cognitive and affective cues that emerge during interaction are not yet leveraged by current approaches. One possible way to complement such feedback is through electroencephalography (EEG), which provides an additional source of information about user state not easily conveyed through language. Because EEG records neural activity with high temporal resolution, it can reflect rapid, pre-verbal reactions to model outputs, such as attention~\cite{makeig2004mining, makeig2002dynamic}, surprise~\cite{garrido2009mismatch}, or cognitive workload~\cite{klimesch1999eeg}. Prior research in brain–computer interfaces~(BCIs) has further shown that EEG correlates with processes including error detection~\cite{holroyd2002neural}, task engagement~\cite{pope1995biocybernetic}, and emotional valence~\cite{davidson1992anterior}, suggesting that it can provide complementary signals for modeling user feedback.

Building on these insights, we explore the use of brain signals (\ie, EEG) as an alternative real-time interface for LLMs. 
By decoding neural responses as users view LLM outputs, EEG signals can provide a indication of satisfaction or dissatisfaction~\cite{ye2022brain, kim2022electroencephalography}. We integrate these signals within the test-time scaling framework~\cite{snellscaling}, which allows a model to allocate additional inference-time computation when feedback suggests dissatisfaction. When EEG patterns indicate that a user may be unsatisfied, the model performs an extra refinement step; otherwise, it returns its response directly to maintain efficiency. This EEG-guided adaptation does not modify model parameters but adjusts inference during test time based on physiological feedback, offering a small step toward exploring how neural information might inform real-time human–AI interaction.

To evaluate the effectiveness of the proposed EEG-based interface pipeline, we conduct a extensive human study, each person completing $130$ image-generation trials while their neural activity is recorded using a $64$-channel EEG system. The experiments show that EEG signals collected during the feedback stage is predictive of user satisfaction~\cite{ye2022brain, kim2022electroencephalography}, suggesting that neural activity carries information relevant to real-time preference inference~\cite{holroyd2002neural}. In addition, the pipeline yields improvements in the alignment between model outputs and user preferences, as well as in the perceived quality of generated content. On average, reported user satisfaction increased compared with baseline settings. We summarize our main contributions as follows:
\begin{itemize}[nosep]
    \item Target people with speech or motor impairments, we are among the first to build Brain–LLM interface as well as a new Dataset, \textbf{BLID}, which is an EEG-based human preference dataset specifically designed for aligning with LLMs. In this dataset, users' neural responses are recorded while they evaluate model-generated outputs, providing a foundation for studying implicit feedback signals (Section~\ref{sec:dataset});
    \item We propose an EEG-guided TTS framework, which leverages EEG signals to adapt LLM outputs in real time, ensuring alignment with user preferences without the need for additional fine-tuning (Section~\ref{sec:method_interface}).
\end{itemize}

\section{Related Work}

\textbf{EEG-based brain–computer interfaces (BCIs)} aim to decode neural activity for applications such as motor imagery classification~\cite{lotte2018review}, emotion recognition~\cite{zheng2018emotionmeter}, and cognitive state monitoring~\cite{roy2019eeg}. Recent advances have expanded EEG decoding beyond traditional single-modality tasks~\cite{wang2022multimodal} toward more ambitious goals, including cross-modal generation~\cite{liu2024eeg2video} and open-vocabulary understanding~\cite{wang2022openvocab}.
In visual perception decoding, early work primarily focused on classifying~\cite{spampinato2017decode} or reconstructing~\cite{miyawaki2008visual} static images and simple visual paradigms. More recent studies have begun addressing richer and more dynamic forms of visual experience. EEG2Video~\cite{liu2024eeg2video}, for example, investigates reconstructing dynamic visual content from EEG signals, while Neuro3D~\cite{guo2025neuro3d} extends this direction toward decoding three-dimensional visual information. At the same time, decoding tasks have broadened from closed visual categories to open semantic systems. EEG2Text~\cite{liu2024eeg2text} explores open-vocabulary EEG-to-text generation, providing a more flexible interface for interpreting user intent. Collectively, these developments move BCIs beyond simple pattern recognition and toward more general content generation and naturalistic forms of human–AI interaction.

\textbf{EEG-based Foundation models.} Recent work in EEG decoding has increasingly focused on narrowing the neural–semantic gap to support more natural forms of human–computer interaction. EEGNet~\cite{lawhern2018eegnet} introduced a compact convolutional architecture that captures spatio-temporal structure in EEG signals, while EEGTransformer~\cite{lee2022eeg} extends this direction by using self-attention to model long-range dependencies. In parallel, self-supervised and large-scale pretraining have emerged as promising strategies for building more generalizable EEG representations. LaBraM~\cite{jiang2024large} provides one of the first large-scale EEG foundation models, learning transferable features across tasks, and ContraWR~\cite{yang2021self} explores contrastive learning for end-to-end feature extraction from raw signals. Of particular relevance to our work are models targeting brain-to-content translation. NeuroLM~\cite{jiang2024neurolm} functions as a specialized language model that decodes semantic intent from EEG, and CBraMod~\cite{wang2024cbramod} aligns such neural representations with textual and generative modalities. Our framework is inspired by these advances and complements them by incorporating EEG feedback into real-time model refinement rather than direct content generation.

\begin{figure*}[htbp]
    \centering
    \includegraphics[width=0.95\textwidth]{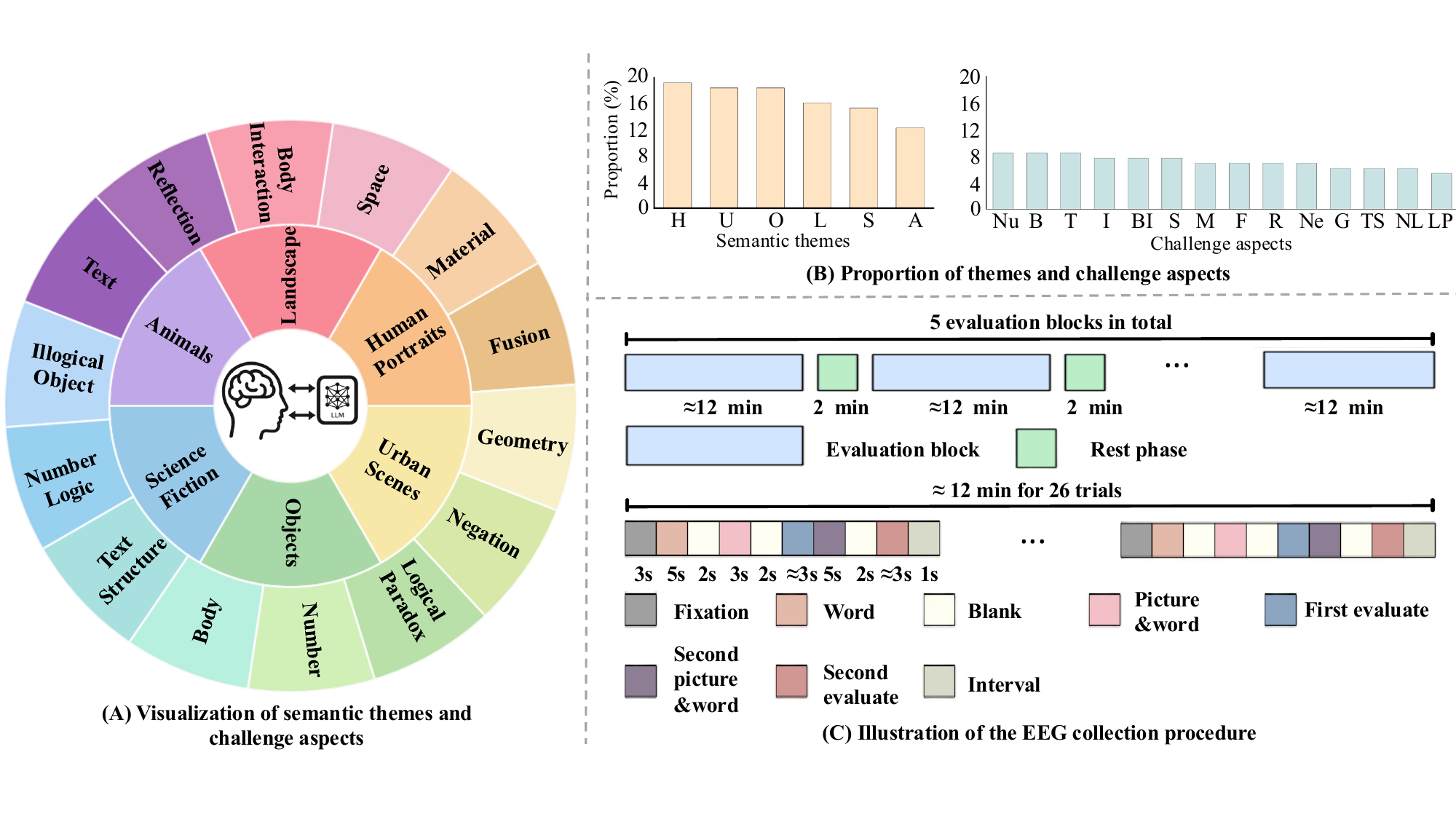}
    \vspace{-0.3in}
    \caption{
    \textbf{Meta-information on the semantic themes, challenge categories, and EEG data collection procedure of BLID.}
\textbf{(A)} Visualization of the six semantic themes and fourteen challenging aspects of text-to-image generation used to construct the stimulus set.
\textbf{(B)} Balanced distribution of prompt–image pairs across six semantic themes and fourteen challenge categories, ensuring diverse coverage of both objective constraints and subjective nuances.
\textbf{(C)} Illustration of the EEG collection procedure. The experiment consisted of five evaluation blocks ($12$ minutes each). Each block contained $26$ trials, and each trial lasted approximately $28$ seconds, including stimulus presentation, evaluation periods, and inter-trial intervals.
}
    \label{fig:dataset}
    \vspace{-0.5em}
\end{figure*}

\textbf{Test-Time scaling (TTS)} enhances model performance by allocating additional computation during inference rather than relying on a single forward pass~\cite{devlin2019bert, houlsby2019parameter, sun2020ttt, shanmugam2020tta}. Traditional approaches typically depend on offline training~\cite{devlin2019bert} or static fine-tuning~\cite{houlsby2019parameter}, which limits adaptability to new inputs or user preferences. More recent test-time adaptation methods, such as test-time training~\cite{sun2020ttt}, test-time augmentation~\cite{shanmugam2020tta}, and online update strategies like TENT~\cite{wang2021tent}, demonstrate that adjusting model behavior at inference can improve robustness and alignment without modifying core model weights.

In this work, we do not propose a new TTS algorithm~\cite{snellscaling}. Instead, we adopt the TTS perspective to build a feedback-driven inference loop: LLM outputs are refined iteratively based on EEG feedback collected after the user views each result. This allows the model to decide, at inference time, whether additional refinement is needed to better align with the user’s preferences.

\begin{figure*}[htbp]
    \centering
    \includegraphics[width=0.95\textwidth]{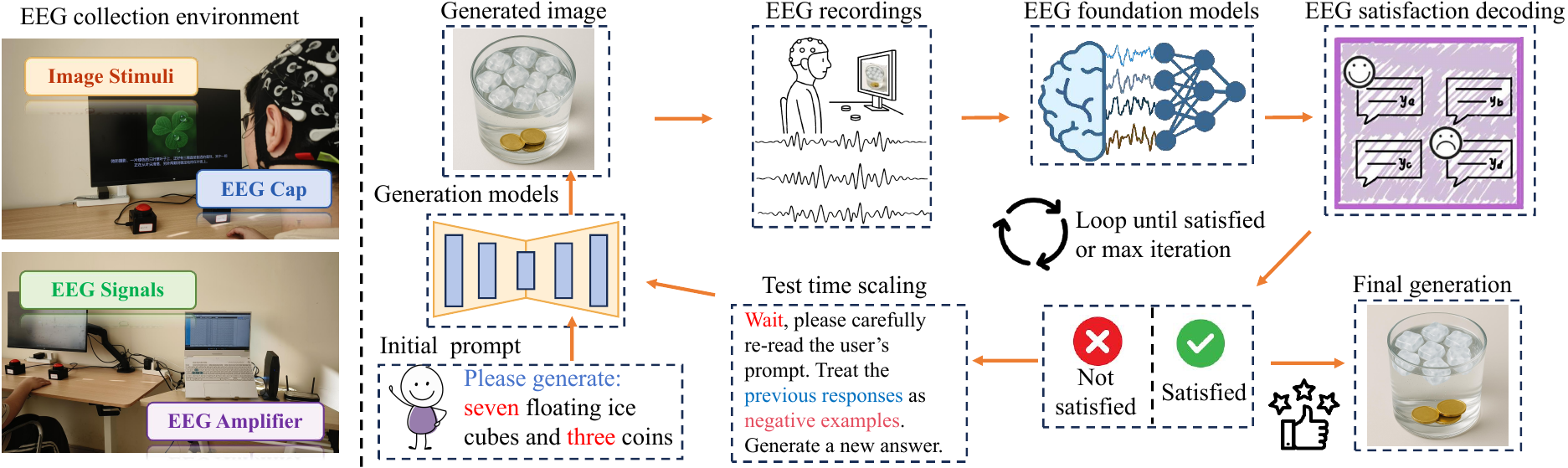}
    \vspace{-0.5em}
    \caption{
    \textbf{Overview of the Brain-LLM interface.} \textbf{Left:} Participants view image stimuli while wearing a 64-channel EEG cap; \textbf{Right:} the signals are recorded and amplified to train an EEG foundation model that predicts user satisfaction. During inference, a generation model produces an initial answer, and the EEG model monitors the user’s neural response, classifying each trial as ``Satisfied'' or ``Not satisfied''. If not satisfied, the system triggers test time scaling step and loops until the user is satisfied or a maximum number of iterations is reached.
    }
    \label{fig:pipeline}
    \vspace{-0.5em}
\end{figure*}

\section{Brain-LLM Interface Dataset}
\label{sec:dataset}
We collect a new dataset, the Brain–LLM Interface Dataset (\textbf{BLID}), to investigate whether (i) EEG signals recorded while users view generated outputs reflect their satisfaction with the generations and (ii) such signals can help improve LLM alignment.

\subsection{Participants}
Ten healthy participants (mean age: 21.5 years; 4 females) with normal or corrected-to-normal vision were recruited. Each person completing 130 image-generation trials while their neural activity is recorded. Thus our dataset contains 1,300 trials. All participants provided written informed consent and paid with salary above minimum level, and the study was approved by the institutional ethics committee.

\subsection{Stimuli Generation}
We prompted GPT-4o to generate 450 textual prompts covering six semantic themes and 14 challenge aspects of text-to-image generation (Figure~\ref{fig:dataset} A). To ensure representativeness and mitigate bias, we employed a \textbf{stratified sampling strategy} to select 130 prompts. This selection ensured a uniform distribution across the 14 challenge categories (see Figure~\ref{fig:dataset}B) while balancing \textbf{objective constraints} (\eg, numeracy) with \textbf{subjective nuances} (\eg, aesthetics). This design captures both gross errors and subtle misalignments typical of real-world interaction. We controlled for prompt complexity to ensure neural responses were driven by satisfaction rather than difficulty.

\begin{algorithm}[t]
   \caption{Brain-LLM Interface for Test-time Scaling}
   \label{alg:eeg_tts}
\begin{algorithmic}[1]
   \STATE {\bfseries Input:} Initial prompt $p_0$, max iterations $T_{\max}$
   \STATE {\bfseries Initialize:} Image generator $G$, LMM operator $R$, EEG decoder $D_\phi$
   \STATE {\bfseries Initialize:} Iteration $t \leftarrow 0$, current prompt $p_t \leftarrow p_0$
   
   \STATE $y_t \leftarrow G(p_t)$ \hfill $\triangleright$ Generate initial image
   
   \WHILE{$t < T_{\max}$}
       \STATE Acquire EEG signal $e_t$ while user views $y_t$
       \STATE $p_t^{\text{score}} \leftarrow D_\phi(e_t)$ 
       \hfill $\triangleright$ Decode satisfaction probability
       
       \IF{$p_t^{\text{score}} \ge \tau_{\mathrm{accept}}$}
           \STATE \textbf{return} $y_t$ 
           \hfill $\triangleright$ Accept the image and terminate
           
       \ELSE
           \STATE $p_{t+1} \leftarrow R(p_0, y_t, p_t)$ 
           \hfill $\triangleright$ Refine prompt
           \STATE $y_{t+1} \leftarrow G(p_{t+1})$
           \STATE $y_t \leftarrow y_{t+1},\; p_t \leftarrow p_{t+1}$
       \ENDIF
       
       \STATE $t \leftarrow t + 1$
   \ENDWHILE
   
   \STATE \textbf{return} $y_t$ 
   \hfill $\triangleright$ Terminate after reaching max iterations or satisfaction
\end{algorithmic}
\end{algorithm}

\subsection{EEG Data Collection}
Data were recorded in a controlled laboratory environment using a 64-channel Neuracle system (active AgCl electrodes, 10–20 system) at a sampling rate of 1,000 Hz. The session lasted approximately 1.5 hours.

\textbf{Procedure.} As illustrated in Figure~\ref{fig:dataset}C, participants evaluated unique prompt–image pairs. Each trial consisted of an initial evaluation and a consistency check. The trial began with a fixation cross (3 s) and the text prompt (5s), followed by the prompt–image pair (3s). Participants then rated the image as ``Satisfied'' or ``Unsatisfied''. A re-evaluation phase followed immediately; trials with inconsistent ratings between the two phases were excluded to ensure label stability. This exclusion is critical because inconsistent responses imply cognitive ambiguity, which fails to elicit the distinct error-processing signals (\eg, ERN) necessary for reliable decoding~\cite{chandrasekar2024machine}. A 1s interval separated consecutive trials. We additionally ensured coverage of both common generation successes and diverse failure patterns to avoid over-representing any single error type.

The experiment comprised five blocks of 26 trials, with the trial order randomized. Short rest periods of approximately 3 minutes were provided between blocks to reduce fatigue and maintain attention. After each rest period, participants reported their self-perceived attention level on a five-point scale (1 indicates very sleepy, 5 suggests highly focused). The results showed an average self-reported attention level of 4.07 across all participants and blocks, indicating that participants were generally able to maintain sufficient concentration for high-quality EEG recording.

\textbf{Data Extraction.} EEG data were extracted from a reaction-time–aligned window spanning $[-2.3, -0.3]$ seconds relative to the button press. This interval isolates preference-related neural activity by excluding motor execution artifacts~\cite{ward2003age}. In total, 100 unique prompt–image pairs were retained, balanced between the ``Satisfied'' and ``Unsatisfied'' conditions.

\section{Method}
Our goal is to incorporate a user’s internal preference signal into the generative loop. Direct EEG-to-text decoding, however, presents clear challenges: decoding arbitrary, open-ended thoughts is not yet reliable for practical use.

To mitigate this issue, we adopt a more coarse-grained and robust strategy. Instead of attempting to decode what \textit{should} be changed in the output, we focus on a binary signal indicating \textit{whether} the user is satisfied or unsatisfied. This simpler signal offers a more reliable basis for guiding the model on whether to accept the current output or perform an additional refinement step. The resulting ``generate–check–refine'' loop aligns naturally with the test-time scaling framework~\cite{muennighoff2025s1, snellscaling}.

\subsection{Recap of Test-Time Scaling}
Test-time scaling (TTS)~\cite{snellscaling}  improves a \emph{frozen} model by allocating extra computation at inference rather than relying on a single forward pass. Common strategies include generating multiple candidates~\cite{madaan2023self}, iteratively refining outputs, and using verifiers or reward models to guide selection~\cite{locey2016scaling}.

We adopt an \emph{iterative revision} view~\cite{li2025testtime}. Let the initial prompt be \(p_0\) and the draft sequence be \(\{y_t\}_{t=0}^{T}\). The model $M_\theta$ first produces a draft:
\[
y_0 = M_\theta(p_0),
\]
and then updates it within the same inference session through a refinement operator \(R\):
\[
y_{t+1} = R\!\left(y_t,\,p_0\right), \quad t=0,1,\dots
\]
The process terminates when a stopping condition is reached, such as exceeding a verification score threshold $\tau$ or hitting a predefined step limit $T_{\max}$:
\[
\text{stop if } \ g(y_t)\ge \tau \ \ \text{or}\ \ t\ge T_{\max},
\]
where \(g(y_t)\) is a verification function that outputs a scalar score representing the quality of the draft \(y_t\).This ``draft–check–revise'' loop can be described as a self-revision method that gradually improves output quality across multiple inference steps. 

In many TTS systems, the decision to continue revising depends on \emph{model-internal} signals, such as verifier scores or heuristic confidence estimates~\cite{li2025testtime}. In the Section~\ref{sec:method_interface}, we retain this iterative-revision framework but replace the decision signal with a preference score \emph{decoded from EEG}, allowing extra computation only when neural evidence indicates user dissatisfaction.

\begin{table*}[t]
\centering
\caption{\textbf{Performance of EEG satisfaction classification.} We compare two categories of models: (1) handcrafted feature–based classifiers (LDA, Logistic Regression, SVM) and (2) end-to-end neural models fine-tuned on raw EEG (EEGNet, EEG-Conformer,CBraMod,LaBraM). Metrics include Accuracy, F1, AUROC, Recall, G-Mean, and MCC. Results show that EEG signals carry information relevant to distinguishing ``\textit{Satisfied}'' \emph{vs.} ``\textit{Unsatisfied}'' states, with LaBraM achieving the highest overall performance.}
\label{tab:model_performance}

% 不要使用 resizebox，改用 \small 或 \footnotesize
\small 
\setlength{\tabcolsep}{4pt} % 稍微调整列间距以适应页面宽度

\begin{tabular}{
  l 
  l 
  S[table-format=2.2(3)] % 格式：2位整数.2位小数(误差)
  S[table-format=2.2(3)] 
  S[table-format=2.2(3)] 
  S[table-format=2.2(3)] 
  S[table-format=2.2(3)] 
  S[table-format=2.2(3)]
}
\toprule
\textbf{Category} & \textbf{Methods} & {\textbf{Accuracy}} & {\textbf{F1-Score}} & {\textbf{AUROC}} & {\textbf{Recall}} & {\textbf{G-mean}} & {\textbf{MCC}} \\
\midrule
\multirow{3}{*}{Hand-crafted} 
 & LDA       & 70.18 \pm 1.65 & 70.72 \pm 2.53 & 74.26 \pm 2.81 & 71.45 \pm 5.40 & 70.01 \pm 1.51 & 40.50 \pm 3.37 \\
 & LR   & 72.32 \pm 2.86 & 72.06 \pm 3.68 & 79.32 \pm 3.09 & 72.05 \pm 6.62 & 72.16 \pm 2.93 & 44.85 \pm 5.69 \\
 & SVM           & 77.85 \pm 2.06 & 77.41 \pm 2.46 & 86.58 \pm 3.22 & 76.67 \pm 4.59 & 77.75 \pm 2.11 & 55.83 \pm 4.06 \\
\midrule
\multirow{4}{*}{End-to-end} 
 & EEG-Conformer & 68.00 \pm 1.77 & 68.18 \pm 1.67 & 72.85 \pm 3.27 & 68.60 \pm 2.88 & 67.94 \pm 1.75 & 36.06 \pm 3.56 \\
 & EEGNet  & 70.20 \pm 1.75 & 69.81 \pm 2.10 & 76.20 \pm 1.67 & 69.00 \pm 3.39 & 70.15 \pm 1.77 & 40.45 \pm 3.49 \\
 & CBraMod   & 77.10 \pm 2.01 & 75.88 \pm 2.38 & 84.78 \pm 1.79 & 72.77 \pm 5.43 & 76.80 \pm 2.00 & 54.59 \pm 4.11 \\ 
 & LaBraM     &  80.68 \pm 2.75 &  80.74 \pm 2.51 &  87.91 \pm 2.83 &  80.97 \pm 3.11 &  80.64 \pm 2.76 &  61.46 \pm 5.50 \\
\bottomrule
\end{tabular}
\end{table*}

\subsection{The Brain-LLM Interface for TTS}
\label{sec:method_interface}

Shown in Figure~\ref{fig:pipeline}, we introduce an EEG-gated interface that operationalizes the iterative refinement paradigm of Test-Time Scaling (TTS). It substitutes model-internal heuristics with direct neural feedback, creating a brain-in-the-loop system that dynamically allocates computational resources. The interface triggers a refinement step only upon detection of user dissatisfaction inferred from their EEG signals. 

The architecture consists of three core modules: a neural preference decoder, an EEG-gated policy, and a large multimodal model (LMM) operator. The entire closed-loop process is formalized in Algorithm~\ref{alg:eeg_tts}. 

\textbf{Neural preference decoder} \(D_\phi\) maps EEG activity to a user’s preference state. It takes a segment of preprocessed EEG data \(e_t \in \mathbb{R}^{C \times T}\) as input and outputs a scalar satisfaction probability \(p_t^{\text{score}} \in [0, 1]\). We formulate this as a binary classification problem and train \(D_\phi\) on the BLID dataset to distinguish neural patterns corresponding to \textit{Satisfied} versus \textit{unsatisfied} evaluations. 

During inference, we convert $p_t^{\text{score}}$ into a binary decision using a fixed acceptance threshold $\tau_{\mathrm{accept}} = 0.5$: responses with $p_t^{\text{score}} \ge 0.5$ are treated as \textit{satisfied} and accepted, while those with $p_t^{\text{score}} < 0.5$ are treated as \textit{unsatisfied} and trigger a refinement step. No subject-specific threshold tuning is performed.

\textbf{LLM TTS for generation refinement via critique and rewrite.} When refinement is triggered (Line 11 in Algorithm~\ref{alg:eeg_tts}), an operator \(R\) produces an improved prompt. This operator is implemented using a frozen, pretrained LMM without any additional fine-tuning. The LMM is prompted to perform a \textit{critique-and-rewrite} procedure: given the original prompt \(p_0\), the generated output \(y_t\), and the intermediate prompt \(p_t\), it generates a revised prompt \(p_{t+1}\) intended to elicit a better response. This prompt-engineering strategy keeps the LMM fixed, ensuring that performance improvements arise solely from the EEG-gated refinement process.

\section{Experiment}
We design three experiments to evaluate the feasibility of using EEG as feedback for LLMs: first, we verify the decodability of user satisfaction; second, we confirm via spatial distribution analysis that the signals originate from cognitive evaluation rather than motor artifacts; and finally, we evaluate the effectiveness of EEG-gated refinement in improving preference alignment within a TTS framework.

\subsection{EEG-based Satisfaction Classification}
To assess whether user preference (Satisfied vs. Unsatisfied) can be reliably decoded from EEG, we conducted a benchmark experiment on the BLID dataset. We evaluate two categories of models: (1) handcrafted feature-based classifiers, including Linear Discriminant Analysis (LDA)~\cite{subasi2010eeg}, Logistic Regression (LR)~\cite{tomioka2006logistic}, and Support Vector Machines (SVM)~\cite{sha2020knn}; and (2) end-to-end neural models fine-tuned on raw EEG, including EEGNet~\cite{lawhern2018eegnet}, EEG-Conformer~\cite{song2022eeg}, CBraMod~\cite{wang2024cbramod}, and LaBraM~\cite{jiang2024large}.

\textbf{Satisfaction can be reliably decoded.} Table~\ref{tab:model_performance} confirms the feasibility of EEG-based preference decoding. Among all models, the large-scale pretrained LaBraM achieves the best performance (80.68\% accuracy, 87.91\% AUROC). Notably, the SVM classifier (77.85\%) outperforms non-pretrained deep architectures like EEGNet (70.20\%), suggesting that traditional deep models struggle to learn robust representations with limited data. In contrast, LaBraM achieves the highest performance by transferring representations learned from large-scale EEG corpora, effectively overcoming challenges related to data scarcity, low signal-to-noise ratios, and inter-subject variability.

\begin{figure}[t]
\centering
\includegraphics[width=\linewidth]{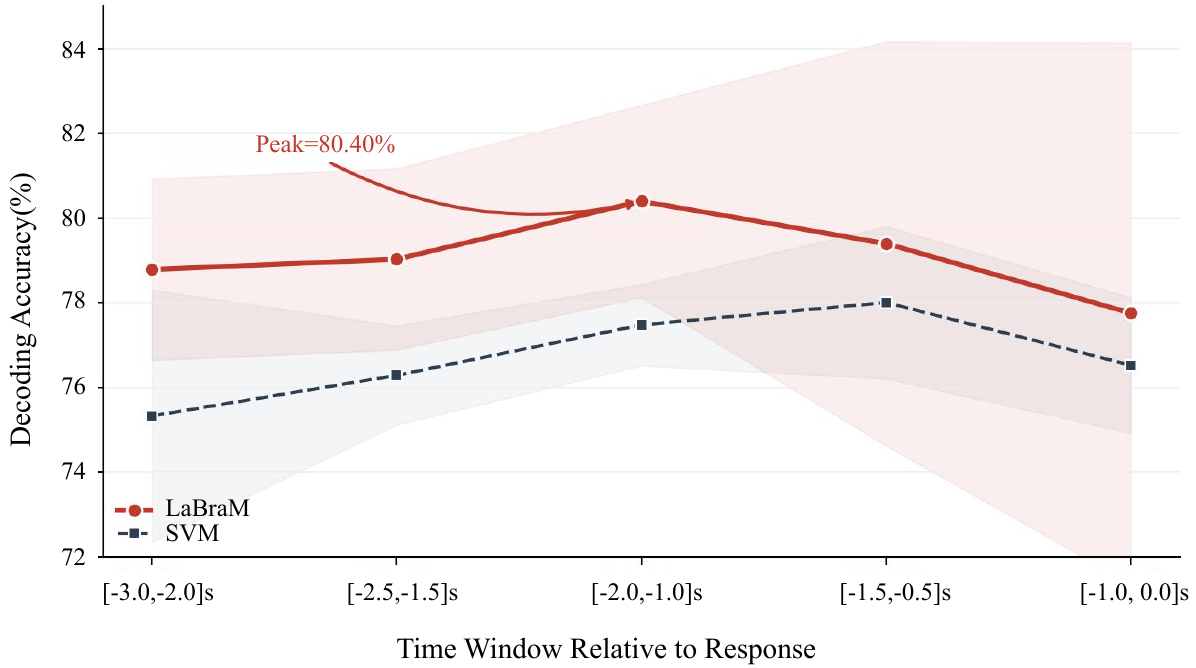}
\caption{\textbf{Time-resolved decoding analysis of user satisfaction.} We evaluate the decoding accuracy of representative hand-crafted (SVM) and end-to-end (LaBraM) models using a sliding window approach relative to the button press (0.0s). The shaded regions represent the standard deviation across participants.}
\label{fig:time_window}
\end{figure}

\begin{figure}
\centering
\includegraphics[width=0.46\textwidth]{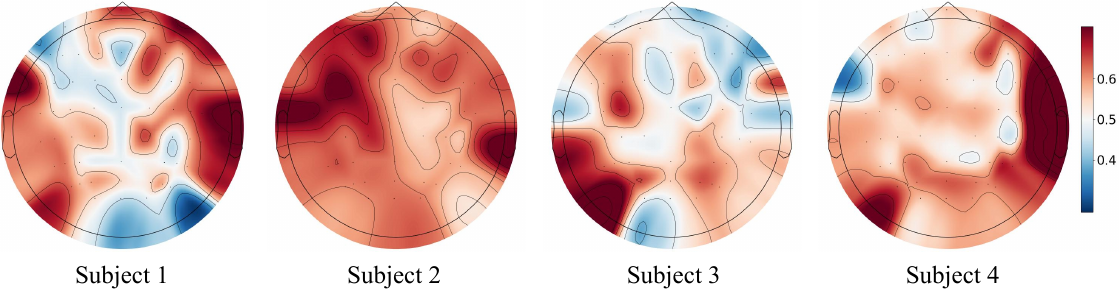}
% \vspace{0.4in}
\caption{\textbf{Topographical maps of single-channel classification accuracy for four participants.} Warm colors indicate electrodes with higher decoding accuracy, and cool colors indicate lower accuracy. While individual variability is present, several regions show consistently higher performance across participants.}
\vspace{0.0em}
\label{fig:brain1}
\end{figure}

\paragraph{Temporal dynamics confirm cognitive origin.}
We further analyzed the time course of decoding accuracy to distinguish preference evaluation from motor execution. As shown in Figure~\ref{fig:time_window}, performance peaks in the $[-2.0, -1.0]$s interval (LaBraM: 80.40\%), significantly preceding the physical response. This suggests that the internal evaluation is finalized well before motor execution. Conversely, as the window approaches the button press ($[-1.0, 0.0]$s), accuracy declines and variance increases, indicating interference from motor preparation noise. This rise in variance near the response point reflects the transition from purely evaluative neural states to complex, participant-specific motor planning rhythms. This trajectory confirms that the decoded signals reflect internal cognitive evaluation rather than motor artifacts, aligning with established neuroscientific evidence that outcome evaluation is a rapid, pre-reflexive process initiated in the medial frontal cortex~\cite{gehring2002medial,chandrasekar2024machine}. Crucially, this temporal decoupling ensures that the preference feedback integrated into our test-time scaling framework is derived from high-level cognitive appraisal rather than the low-level physiological noise of physical interaction.

\begin{table*}[t]
\centering
\footnotesize
\caption{\textbf{Effectiveness of the Brain-LLM interface.}
We apply the framework to a range of state-of-the-art text-to-image models, including both open-source and proprietary systems. Performance is evaluated using 6 metrics: CLIP Score, BLIP Score, Aesthetic Score, VQA Score, the average human rating (Human, 1–5), and the final success rate,which represents the percentage of images the user finds satisfactory after evaluation through The Brain-LLM Interface for TTS. Across all models, the Brain-LLM framework consistently improves generation quality and preference alignment relative to the baseline.}
\label{tab:image_model_comparison}
\setlength{\tabcolsep}{0.8mm}  
% \resizebox{\textwidth}{!}{
\begin{tabular}{llccccccc}
\toprule
\textbf{Category} & \textbf{Model} & \textbf{Condition} & \textbf{CLIP Score} & \textbf{BLIP Score} & \textbf{Aesthetic} & \textbf{VQA Score} & \textbf{Human} & \textbf{Success Rate} \\
\midrule
\multirow{8}{*}{Open Source} 
& \multirow{2}{*}{Stable Diffusion XL } & First Generation & 27.23 & 88.90 & 6.11 & 37.72 & 2.56 & 3.33 \\
& & Brain-LLM & 28.71 & 97.41 & 6.51 & 52.21 & 3.08 & 27.62 \\
\cmidrule(lr){2-9}
& \multirow{2}{*}{Wanx-v1} & First Generation & 25.35 & 84.81 & 6.31 & 41.62 & 2.62 & 6.67 \\
& & Brain-LLM & 27.70 & 88.04 & 6.65 & 51.06 & 2.91 & 17.62 \\
\cmidrule(lr){2-9}
& \multirow{2}{*}{Wan2.2-t2i-flash} & First Generation & 26.34 & 96.57 & 6.21 & 35.48 & 3.34 & 33.33 \\
& & Brain-LLM & 27.56 & 97.56 & 6.43 & 45.33 & 3.81 & 65.24 \\
\cmidrule(lr){2-9}
& \multirow{2}{*}{Qwen-image-plus} & First Generation & 26.14 & 92.25 & 6.02 & 45.27 & 3.31 & 39.05 \\
& & Brain-LLM & 28.05 & 95.59 &  6.57 & 50.98 & 4.01 & 80.48 \\
\midrule
\multirow{6}{*}{Closed Source} 
& \multirow{2}{*}{GPT-4o Image} & First Generation & 28.96 & 97.37 & 6.22 & 55.14 & 3.67 & 64.76 \\
& & Brain-LLM & 30.27 & 98.44 & 6.37 & 62.92 & 4.16 & 92.38 \\
\cmidrule(lr){2-9}
& \multirow{2}{*}{Doubao-seedream-3-0-t2i} & First Generation & 28.95 & 99.79 & 6.07 & 65.05 & 3.64 & 55.24 \\
& & Brain-LLM & 30.26 & 99.91 & 6.28 & 77.37 & 4.06 & 92.38 \\
\cmidrule(lr){2-9}
& \multirow{2}{*}{Doubao-seedream-4-0} & First Generation & 28.97 & 99.18 & 6.11 & 59.74 & 4.13 & 58.10 \\
& & Brain-LLM & 29.96 & 99.81 & 6.29 & 72.48 & 4.23 & 95.24 \\
\bottomrule
\end{tabular}
% }  
\end{table*}

\begin{figure}[t]
\centering
% 1. 左侧放图片的 minipage，使用 [c] 垂直居中
\begin{minipage}[c]{0.23\textwidth}
    \centering
    \includegraphics[width=\linewidth]{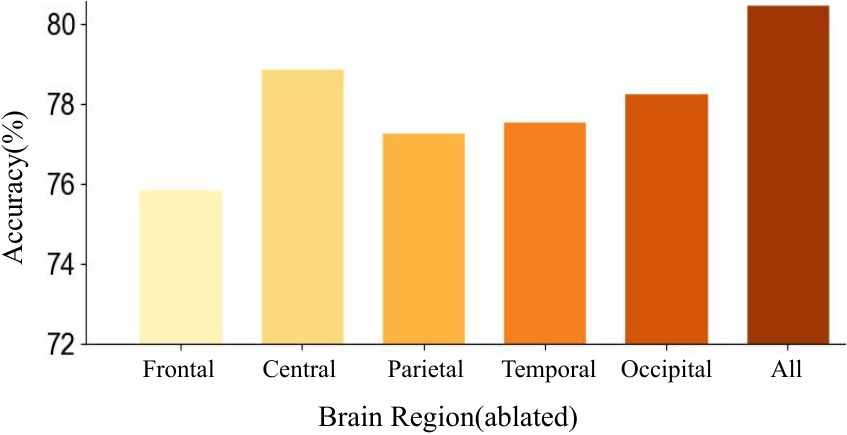}
\end{minipage}%
\hfill % 在中间加弹性空白
% 2. 右侧放 Caption 的 minipage，同样使用 [c]
\begin{minipage}[c]{0.23\textwidth}
    \caption{\textbf{Classification accuracy ablation.} The rightmost bar (''All'') shows the baseline performance using all electrodes. The frontal region contributes most significantly to decoding performance.}
    \label{fig:brain2}
\end{minipage}
% \vspace{-1em} % 减少下方空白
\end{figure}

\subsection{Analysis of Brain Regions}
To identify cortical areas critical for subjective evaluation, we conducted single-channel classification and systematic brain region ablation.

\textbf{Spatial distribution of evaluative signals.} Channel-wise analysis (Figure~\ref{fig:brain1}) shows high decoding accuracies concentrated in \textbf{frontal regions} and \textbf{temporo-occipital areas}, aligning with prefrontal value-based decision-making \cite{rushworth2012valuation}, visual categorization \cite{grill2014functional}, and aesthetic appraisal \cite{chatterjee2014neuroaesthetics}. The frontal involvement likely reflects the cognitive arbitration between model output and user expectations, while posterior activity highlights visual-semantic matching. Critically, the \textbf{central sensorimotor strip}—the hub for motor execution \cite{pfurtscheller1999event, wolpaw2002brain}—exhibits low accuracy. This "silence" validates our pre-response analysis window ($-2.3$s to $-0.3$s) and confirms that the framework successfully isolates purely evaluative neural correlates from motor execution noise and preparation artifacts, ensuring that extracted features are driven by internal appraisal rather than anticipatory motor activity.

\textbf{Functional necessity via ablation.} As shown in Figure~\ref{fig:brain2}, the full-channel model achieves $80.68\%$ accuracy. Removing \textbf{frontal regions} causes the largest drop (to $75.88\%$), reinforcing the prefrontal cortex's dominant role in judgment \cite{miller2001integrative, rushworth2012valuation}. Conversely, removing the \textbf{central region} yields the smallest decrease, proving that motor activity is not a driver for our decoder. Modest reductions in other areas suggest a distributed network integrating attention and visual-semantic processing \cite{corbetta2002control, grill2014functional}. Collectively, this spatial and functional evidence demonstrates that our decoder leverages higher-order signals related to \textbf{evaluative judgment and semantic consistency} rather than low-level motor responses~\cite{miller2001integrative, rushworth2012valuation}, supporting EEG as a \textbf{robust} feedback source for real-time model refinement.

\begin{figure}
\centering
\includegraphics[width=0.46\textwidth]{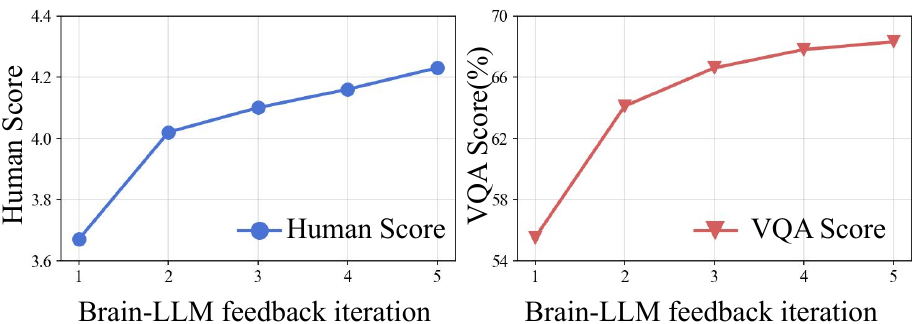}
\caption{\textbf{The effect of TTS on feedback iterations.} This figure shows the change in GPT-4o performance across Human Score and VQA Score as the number of Brain-LLM feedback iterations increases. Both metrics exhibit a steady upward trend, suggesting that iterative EEG-guided refinement can provide incremental improvements to the generated outputs.}
% \vspace{1.0em}
\label{fig:trend}
\end{figure}

\begin{figure*}[htbp]
  \centering
  \includegraphics[width=\linewidth]{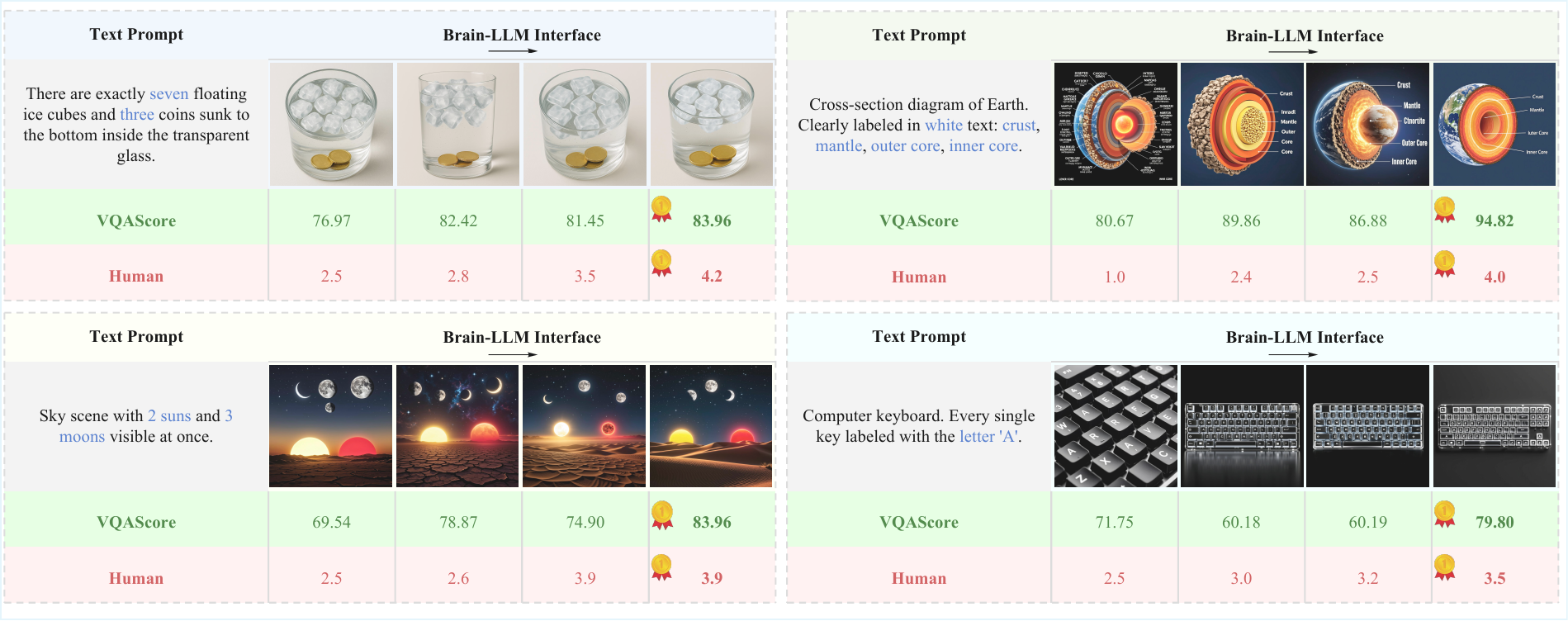}
  \caption{\textbf{Examples of improvement through Brain-LLM feedback in image generation.} This figure illustrates the gradual improvements achieved through Brain-LLM feedback across several examples. For each prompt, we compare VQA Score and Human Score before and after iterative feedback. Images on the left show the initial generations, while those on the right show the outputs after Brain-LLM refinement. As feedback iterations increase, the generated images show clearer details and better adherence to the prompt, suggesting that EEG-guided refinement can contribute to incremental improvements in model performance.}
  \label{fig:example_improvement}
  % \vspace{-0.5em}
\end{figure*}

\subsection{Model Alignment with Human Preference}
In this section, we evaluate the effectiveness of our Brain-LLM interface in improving the alignment of text-to-image models with human preferences. We apply the framework to a range of state-of-the-art models, including open-source systems such as Stable Diffusion XL~\cite{podell2023sdxl}, Wanx-v1~\cite{ma2024visual}, Wan2.2-t2i-flash~\cite{wan2025wan}, and Qwen-image-plus~\cite{wang2024qwen2}, as well as proprietary models such as GPT-4o Image~\cite{hurst2024gpt}, Doubao-seedream-3-0-t2i~\cite{gao2025seedream}, and Doubao-seedream-4-0~\cite{seedream2025seedream}. We evaluate performance using six metrics: CLIP Score~\cite{radford2021learning}, BLIP Score~\cite{li2022blip}, Aesthetic Score~\cite{murray2012ava}, VQA Score~\cite{lin2024evaluating}, the average human rating (Human) from 1 to 5, and the Success Rate, which represents the percentage of images the user finds satisfactory after evaluation through The Brain-LLM Interface for TTS.

\paragraph{Brain-LLM consistently improves alignment and generation quality over baseline.} As shown in Table~\ref{tab:image_model_comparison}, across all seven generative models and six evaluation metrics, the Brain-LLM framework yields consistent improvements. The most notable gains appear in the two metrics most sensitive to human judgment and constraint satisfaction: Human Score and VQA Score. This suggests that EEG-gated refinement may help the model better capture user preferences and adhere more closely to prompt specifications. 

To better understand how the framework achieves these gains, we examined how performance metrics evolve across Brain-LLM refinement iterations. As shown in Figure~\ref{fig:trend}, both Human Score and VQA Score increase steadily with each feedback loop. This pattern suggests that EEG-guided refinement provides incremental improvements rather than random variation, gradually correcting earlier errors and moving the output toward the user's intended specification. Crucially, the steady upward trend across five iterations observed in Figure~\ref{fig:trend} demonstrates that EEG feedback provides a consistent optimization gradient rather than a simple random ``restart''. This sustained improvement suggests that the decoded preference signal remains highly informative even as output quality increases, effectively guiding the model to address subtle misalignments that are typically missed in unguided or zero-shot generation. Additionally, it is also important to verify that these improvements manifest in challenging cases. Figure~\ref{fig:example_improvement} presents representative examples from our stress-test prompts. For example, with the ``\textit{seven ice cubes and three coins}'' prompt, the initial output fails to meet the counting and spatial requirements; after EEG-guided refinements, the final image satisfies both constraints. For ``\textit{cross-section of Earth}'', the clarity of textual labels improves progressively across iterations, yielding a more interpretable and accurate depiction.

\section{Discussion and Future Work}
Looking ahead, our study represents a foundational step toward neural-feedback guided language model adaptation. While the current EEG feedback is simplified to binary signals, this robust ``binary gate'' provides a high-reliability confirmation channel essential for users with severe motor impairments, for whom complex language communication is often challenging. Future research could leverage multi-dimensional neural markers—such as cognitive load, confidence, or affective valence—to provide a more nuanced ``voice'' for those who cannot provide explicit linguistic corrections, thereby refining model behavior without the need for motor-based input.

% \section{Future Work}
Beyond discrete satisfaction signals, our ultimate vision is to establish a unified EEG-text-LLM interaction paradigm that restores interactive agency to individuals with profound physical constraints. We aim to move toward direct semantic synchronization, where EEG acts not merely as a validation tool, but as a continuous stream of cognitive context. This would allow the LLM to perceive and proactively adapt to the user’s internal state as intent emerges. Such an advancement would evolve the interface from a corrective loop into a seamless, brain-driven collaborator, effectively bypassing physical limitations to enable autonomous human-AI synergy.

\section{Conclusion}
This work introduces an EEG-guided test-time scaling framework that adapts LLM outputs using real-time neural feedback. By decoding brain activity, our interface provides an alternative interaction channel that connects user intent directly to model generation, reducing the reliance on explicit motor or linguistic input. Experiments confirm that EEG signals carry decodable satisfaction information, which significantly improves preference alignment when integrated into iterative refinement. These findings demonstrate that neural feedback can inform adaptive inference, paving the way for more inclusive and accessible AI systems. While currently focused on binary signals, future extensions will explore richer neural states to enable more flexible guidance, eventually bridging the gap between human neural activity and generative intelligence, and restoring interactive agency to those previously limited by physical constraints.

\section*{Impact Statement}

This research introduces a Brain-LLM interface that leverages non-invasive EEG signals to align large language model outputs with implicit human preferences. The potential societal impacts of this work are significant and multifaceted:

\begin{itemize}[leftmargin=*, nosep]
    \item \textbf{accessibility and Inclusion:} Our framework provides a promising interaction channel for individuals with severe speech or motor impairments (\eg, ALS or motor neuron diseases). By utilizing neural signals as an implicit feedback source to guide real-time model refinement, we take a step toward making generative AI more accessible to populations who may struggle with traditional linguistic or motor-based interfaces.
    
    \item \textbf{privacy and Neuro-ethics:} As brain-computer interfaces (BCIs) become more integrated with LLMs, the protection of "mental privacy" is of paramount importance. Our study focuses on task-specific, coarse-grained satisfaction signals. However, we emphasize that future developments must prioritize rigorous data encryption, informed consent, and safeguards against the unauthorized decoding of private cognitive or affective states.
    
    \item \textbf{human-AI Collaboration:} This work advances human-centered machine learning by providing a direct link between internal cognitive states and AI generation. While this enhances personalization, we encourage the community to consider the implications of automated preference alignment on human agency and to ensure that such systems are used to augment, rather than replace, deliberate human decision-making.
\end{itemize}

Overall, this paper presents work whose goal is to advance the fields of Machine Learning and BCI. We believe the societal consequences are largely positive for human-AI synergy, provided that neuro-ethical standards are strictly maintained as the technology evolves.

\section*{Acknowledgments}
This work was supported by the Key Research and Development Program of Shandong Province, China (2025CXGC010901), and the Shandong Province Overseas Young Talents Program.

\bibliographystyle{icml2026}

%%%%%%%%%%%%%%%%%%%%%%%%%%%%%%%%%%%%%%%%%%%%%%%%%%%%%%%%%%%%%%%%%%%%%%%%%%%%%%%
%%%%%%%%%%%%%%%%%%%%%%%%%%%%%%%%%%%%%%%%%%%%%%%%%%%%%%%%%%%%%%%%%%%%%%%%%%%%%%%
% APPENDIX
%%%%%%%%%%%%%%%%%%%%%%%%%%%%%%%%%%%%%%%%%%%%%%%%%%%%%%%%%%%%%%%%%%%%%%%%%%%%%%%
%%%%%%%%%%%%%%%%%%%%%%%%%%%%%%%%%%%%%%%%%%%%%%%%%%%%%%%%%%%%%%%%%%%%%%%%%%%%%%%
\newpage
\appendix
\onecolumn

% --- 补充材料大标题 ---
% Adapted for ICML one-column appendix style
\begin{center}
    \Large \textbf{EEG-Based Brain-LLM Interface for Human Preference Aligned Generation} \\
    \vspace{0.5em}
    \large \textbf{Supplementary Material} \\
    \vspace{1.0em}
\end{center}
% --------------------

\section{EEG Data Preprocessing}

In this section, we detail the EEG preprocessing pipelines. To ensure a fair comparison while still allowing each backbone to operate under its most suitable conditions, we followed preprocessing strategies aligned with their respective official implementations or commonly adopted benchmark configurations.

\vspace{0.5em}
\noindent\textbf{General procedures.}
EEG was recorded using a 64-channel Neuracle NeuSen W system. 
We discarded five non-EEG channels (\texttt{VEOL}, \texttt{VEOU}, \texttt{HEOL}, \texttt{HEOR}, \texttt{ECG}), resulting in $59$ channels used in all analyses.
A $50$~Hz notch filter was universally applied to mitigate power-line interference. Data were segmented into epochs ranging from $-2.3$~s to $-0.3$~s relative to the response onset markers, \ie, EEG recorded $0.3$~s before the user pressed the button. To ensure a balanced evaluation across all models, we selected at most $50$ samples per class for each subject using \textbf{random sampling}.

\vspace{0.5em}
\noindent\textbf{End-to-end models.}
For \textbf{LaBraM}~\cite{jiang2024large} and \textbf{EEGNet}~\cite{lawhern2018eegnet}, we adopted a continuous-stage strategy with a $0.1$--$75.0$~Hz bandpass filter to capture broad spectral information. 
LaBraM was subsequently downsampled to $200$~Hz with independent channel-wise $z$-score normalization, following its published configuration.
In contrast, \textbf{EEGNet} operated on the native $1000$~Hz sampling rate of our acquisition system without additional downsampling. This choice is consistent with the temporal-convolutional design of EEGNet, which is typically configured relative to the sampling rate, and avoids under-tuning its temporal resolution. Class imbalance during training was handled by a \textit{WeightedRandomSampler}.
For \textbf{EEG-Conformer}~\cite{song2022eeg}, we followed a segment-first pipeline with a $1.0$--$40.0$~Hz bandpass filter and downsampling to $256$~Hz, and applied fold-wise fitting of the scaler to strictly prevent data leakage.

\vspace{0.5em}
\noindent\textbf{Hand-crafted feature baselines.}
For SVM~\cite{sha2020knn}, LR~\cite{tomioka2006logistic}, and LDA~\cite{subasi2010eeg}, preprocessing focused on the $8.0$--$45.0$~Hz frequency band to specifically extract oscillatory features (alpha, beta, and low gamma). The signals were downsampled to $200$~Hz and baseline-corrected (using a $0.3$~s pre-response interval). Covariance matrices were computed and projected via \textit{tangent space mapping} to align non-Euclidean geometric features before classification.

\section{Experimental Setup and Evaluation Metrics}

In this section, we provide detailed configurations for model implementation and define the quantitative metrics used to assess alignment and generation quality.

\subsection{Implementation Details}

\noindent\textbf{EEG Decoding Models.} 
All EEG classifiers were implemented in PyTorch using the AdamW optimizer. For \textbf{LaBraM}, we fine-tuned the classification head of the pretrained checkpoint (TUH EEG Corpus) with a learning rate of $1e\text{-}4$, a batch size of $16$, and trained for $30$ epochs. For \textbf{EEGNet} and \textbf{EEG-Conformer}, we trained the models from scratch with a higher learning rate of $1e\text{-}3$ and a batch size of $16$, running for $30$ epochs to ensure convergence.

\noindent\textbf{LMM Refinement Operator.} 
For the closed-loop generation, we utilized state-of-the-art vision-language models (\eg, GPT-4o, Doubao-Vision) as the operator $R$. To balance creativity and instruction following, the temperature was set to $0.7$ for the critique-and-refine step, and $0.0$ for the evaluation step to ensure deterministic feedback. The maximum number of refinement iterations $T_{\max}$ was fixed at $5$. The system prompt instructed the model to ``\textit{fix defects inconsistent with the target}'' and ``\textit{reinforce critical constraints}''.

\subsection{Evaluation Metrics Details}

We employed a comprehensive suite of automated and human metrics to evaluate the framework.

\noindent\textbf{Automated Metrics.}
\begin{itemize}[leftmargin=*]
    \item \textbf{CLIP Score}~\cite{radford2021learning}: We follow the standard CLIP-Score implementation and compute this metric using the \texttt{openai/clip-vit-large-patch14} backbone, which embeds both the generated images and the text prompts into a shared representation space and reports their cosine similarity.
    \item \textbf{Aesthetic Score}~\cite{murray2012ava}: Evaluates visual quality using a linear probe trained on the LAION-Aesthetics dataset, built upon the CLIP ViT-L/14 features.
    \item \textbf{VQA Score}~\cite{lin2024evaluating}: To assess fine-grained constraint satisfaction, we adopt the VQA Score metric implemented with the \texttt{CLIP-FlanT5-XXL} model. For each prompt–image pair, we follow the official yes/no question template that asks whether the image matches the full textual description and we use the model's probability of an affirmative answer as the VQA Score.
    \item \textbf{BLIP Score}~\cite{li2022blip}: Measures the semantic consistency between the generated image and the original prompt using the Image-Text Contrastive (ITC) score from the BLIP model.
\end{itemize}

\noindent\textbf{Human Evaluation.}
We conducted a user study with $5$ participants. For each image, participants provided:
\begin{itemize}[leftmargin=*]
    \item \textbf{Human Rating (1--5):} A Likert scale measuring overall alignment and quality ($1$ = Poor, $5$ = Perfect).
    \item \textbf{Success Rate (\%):} The percentage of trials where the participant explicitly evaluated the final generated image as \textit{Satisfied}. To ensure the accuracy of this metric, we rely solely on the participant's \textbf{manual ground-truth feedback}, independent of the EEG decoder's real-time predictions.
\end{itemize}

\section{Analysis of Individual Differences}

\begin{figure}
\centering
\includegraphics[width=0.46\textwidth]{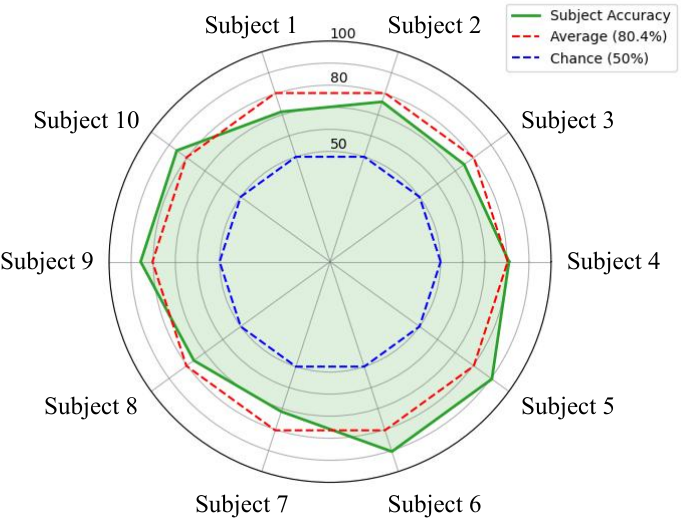}
\caption{\textbf{Radar plot visualizing the decoding performance of ten participants on the binary preference classification task.} The green solid line represents individual subject accuracies, while the red and blue dashed lines indicate the group average (80.4\%) and the chance level (50\%), respectively.}
\vspace{0.0em}
\label{fig:Radar}
\end{figure}

As shown in Fig.~\ref{fig:Radar}, we visualize the decoding performance of all ten participants on the binary preference classification task using a radar plot. All subjects achieve accuracies clearly above the random baseline, indicating that preference-related information is reliably encoded in their EEG signals and can be effectively extracted by our classifier. At the same time, there is noticeable between-subject variability: some participants perform slightly above the group average, whereas others are slightly below, reflecting the widely reported individual differences in EEG recordings. 

Despite this variability, all subjects remain within a practically acceptable performance range, suggesting that the proposed decoder is applicable to heterogeneous user populations. However, the pronounced individual differences also highlight a key challenge for cross-subject EEG-based preference decoding: performance without subject-specific calibration is still limited. Mitigating this cross-subject variability and improving the generalization of the model across users will be an important direction for future work.
\section{Limitations of Base Image Generators under Stress-Test Prompts}

To better understand whether the observed failure cases are specific to a
particular image generator, we additionally evaluated several different
backbones on our high-difficulty prompt set (see
Table~\ref{tab:image_model_comparison}), including both closed-source
models and an open-source diffusion model such as Stable Diffusion XL (SDXL).
We observe that stronger backbones generally yield higher automatic scores
(\eg, CLIP, BLIP, aesthetic, and VQA scores) and noticeably improve the
human-rated success rate. However, even the best-performing backbones still
fail on a non-negligible portion of the stress-test prompts: for example, the
strongest closed-source models reach success rates of only around
$70$--$75\%$, with many failures concentrated on fine-grained counting,
compositional spatial relations, and precise text rendering.

These results suggest that the bottleneck is not solely in the proposed
Brain-LLM interface or the EEG decoder, but also in the current generation
capabilities of base T2I models when faced with highly structured,
constraint-heavy prompts. In other words, simply swapping in a stronger backbone (\eg, GPT-4o Image or Doubao-seedream-4-0) improves the scores but does not fundamentally resolve the hardest cases in our benchmark. We therefore view
our framework as complementary to future advances in image generators:
better backbones are expected to raise the overall performance ceiling,
while EEG-guided test-time scaling can help more reliably steer the
generative process toward user-preferred solutions within that ceiling.

\end{document}